\documentclass{article}

\usepackage{amsmath,amssymb}
\usepackage{stmaryrd}
\usepackage{enumerate}
\usepackage{array}
\usepackage{bbm}
\usepackage{eqparbox}
\usepackage{frame}
\usepackage{float}
\usepackage{algorithm}
\usepackage{color}
\usepackage{url}
\usepackage{epsfig}
\usepackage{multirow}
\usepackage{color}
\usepackage{subfig}
\usepackage{cite}
\usepackage{todonotes}
\usepackage[font=small]{caption}
\usepackage{balance}
\usepackage{graphicx}
\usepackage{adjustbox}
\usepackage{geometry}
\usepackage{spconf}
\newcommand{\beq}{\begin{equation}}
\newcommand{\eeq}{\end{equation}}
\newcommand{\bdm}{\begin{displaymath}}
\newcommand{\edm}{\end{displaymath}}

\newtheorem{definition}{Definition}

\newcommand{\bd}{\begin{definition}}
\newcommand{\ed}{\end{definition}}

\newcommand{\bv}{\begin{vugraph}}
\newcommand{\ev}{\end{vugraph}}
\newcommand{\bi}{\begin{itemize}}
\newcommand{\ei}{\end{itemize}}
\newcommand{\ben}{\begin{enumerate}}
\newcommand{\een}{\end{enumerate}}

\newcommand{\bean}{\begin{eqnarray*} }
\newcommand{\eean}{\end{eqnarray*} }
\newcommand{\bea}{\begin{eqnarray} }
\newcommand{\eea}{\end{eqnarray} }

\newcommand{\ba}{\begin{array} }
\newcommand{\ea}{\end{array} }

\begin{document}
\pagenumbering{gobble}
\title{The Coupled TuFF-BFF Algorithm for Automatic 3D Segmentation of Microglia}
\name{Tiffany Ly$^{\dagger}$, Jeremy Thompson$^{\ddagger}$, Tajie Harris$^{\ddagger}$, and Scott T. Acton$^{\dagger}$, \textit{Fellow, IEEE}}
\address{$^{\dagger}$C.L. Brown Department of Electrical \& Computer Engineering \\ $^\ddagger$ Center for Brain Immunology and Glia, Department of Neuroscience, University of Virginia\\
Charlottesville, Virginia USA}
%{tiffany@virginia.edu} \\ \\
%\date{January 31 2018} 
\maketitle

\begin{abstract}

\end{abstract}

We propose an automatic 3D segmentation algorithm for multiphoton microscopy images of microglia. Our method is capable of segmenting tubular and blob-like structures from noisy images. Current segmentation techniques and software fail to capture the fine processes and soma of the microglia cells, useful for the study of the microglia role in the brain during healthy and diseased states. Our coupled tubularity flow field (TuFF)-blob flow field (BFF) method evolves a level set toward the object boundary using the directional tubularity and blobness measure of 3D images. Our method found a 20$\%$ performance increase against state of the art segmentation methods on a dataset of 3D images of microglia even in images with intensity heterogeneity throughout the object. The coupled TuFF-BFF segmentation results also yielded 40$\%$ improvement in accuracy for the ramification index of the processes, which displays the efficacy of our method. 

\section{Introduction}
\label{sec:intro}
Despite the fact that glia occupy some 80$\%$ of the human brain, the automated segmentation of microglia cells is an open problem. Recent studies in field of neuroscience have shown that studying the morphology of microglia in different scenarios may give significant insight to neurological diseases and brain injury. Microglia are the tissue resident macrophages of the brain parenchyma and have diverse roles in brain development, homeostasis and in injury and disease \cite{colonna2017microglia, schafer2015microglia}. Pioneering in vivo studies demonstrated that microglia processes are constantly in motion even in the healthy brain and were therefore ascribed a surveillant function \cite{nimmerjahn2005resting, davalos2012fibrinogen}. Microglial morphology and behavior are known to be indicative of the physiologic state of the brain and are likely intricately linked with their functions in the healthy brain \cite{perry2010microglia}. The constant motion of microglial processes is postulated to be important for allowing microglia to sense and respond rapidly to their environment including monitoring synaptic activity, sensing invading pathogens and dying cells, and responding to injury \cite{wake2009resting, tremblay2010microglial, davalos2012fibrinogen,madry2017microglial}. During brain injury and disease this constant movement is altered as microglia retract. Their processes take on a more amoeboid morphology, however, little is known about how the decrease in microglia process movement affects their ability to perform their functions in the brain.
%Microglia is an immune cell in the brain that is responsible for tissue repair, phagocytose of dead cells, and other pathological functions \cite{davalos2012fibrinogen}. The long, branch-like processes of these cells is known to survey the brain during these functions \cite{davalos2012fibrinogen, nimmerjahn2005resting}. However, little is known about the role of microglia during brain infection. 

Microglia morphology and behavior are complex and methods of automatic segmentation and analysis of these aspects of microglia are lacking in the field of image processing. Through multiphoton microscopy imaging, it is apparent that the morphology and movement of microglia differs significantly between the brains of healthy mice and the brains of infected mice. Microglia of infected mice have decreased sampling of tissue in part due to the ramification of microglia processes. With high-throughput imaging, we may be able to reveal a quantitative model of these differences. However, there are currently no segmentation tools that are specific to the segmentation of microglia.

\begin{figure}[t!]
	\centering
	\renewcommand{\tabcolsep}{0.05cm}
	\setlength{\belowcaptionskip}{-10pt}
	\begin{adjustbox}{width=\linewidth}
	{
			\begin{tabular}{cc}
				%	\begin{tabular}{@{}p{2cm}p{2cm}p{2cm}p{2cm}p{2cm}p{2cm}p{2cm}p{2cm}p{2cm}p{2cm}@{}}
				\includegraphics[width=.18\linewidth, height = 0.13\linewidth]{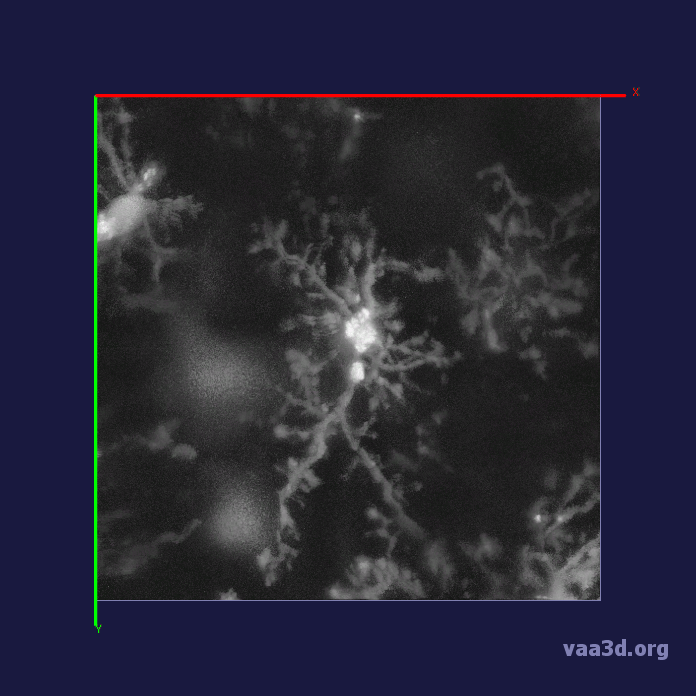}&
				\includegraphics[width=.18\linewidth, height = 0.13\linewidth]{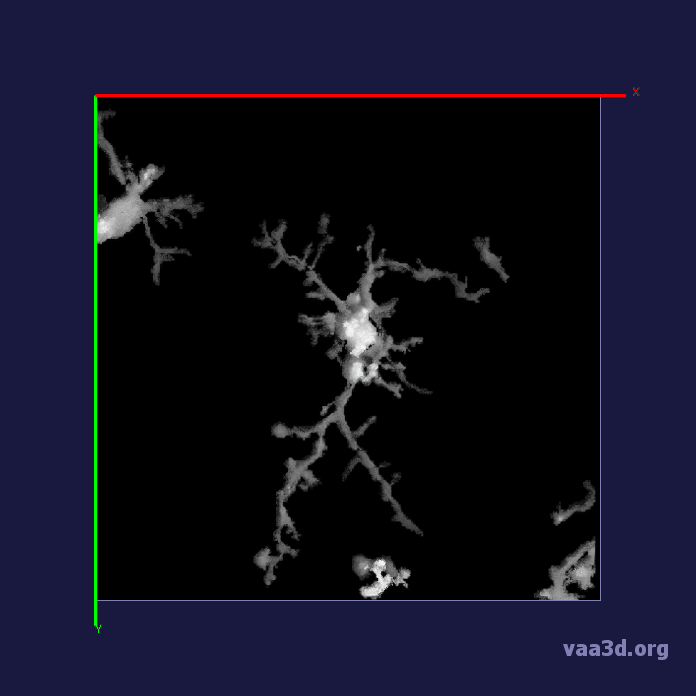}				
			\end{tabular}
		}	
	\end{adjustbox}
%	\vspace{-0.1cm}
	\caption{\small{3D microglia image from multiphoton microscopy. Segmentation of microglia using coupled TuFF-BFF.}}
	
	\vspace{-0.5cm}
	\label{fig:onecell}
\end{figure}

Nimmerjahn \textit{et al.} manually traced the ends of the processes to get a rough estimation of the velocity of length change and drew out microglia for other measurements \cite{nimmerjahn2005resting}. This manual method does not give accurate measurements for the fine processes and is not feasible for high throughput data. Others quantified microglia size and processes movement by thresholding the foreground and background \cite{davalos2005atp}, manually outlining the cell, and manually counting primary branches using ImageJ software (National Institutes of Health) \cite{gyoneva2014systemic}. The most automated segmentation effort for microglia images was reported by Madry \textit{et al.} in which Vaa3D software is used to trace microglia \cite{madry2017microglial}. These methods were usually done in 2D and is not a good fit for images that have high intensity inhomogeneity and background noise. As discussed later in Section \ref{sec: exp}, imaging microglia from healthy and infected mice with multiphoton microscopy result in images with varying intensity contrast throughout the cell which makes it difficult to threshold and separate the object from the background.

%Lit review on something
%background on microglia segmentation/reconstruction... Davalos, Nimmerahn, davolos 2014...
%uses thresholding (atwell... vaa3d)
%background on tracing techniques
In this paper, we propose an automatic method for segmentation of 3D images of microglia. Our method can capture the fine processes and soma in noisy images without prior processing. We compare our method to state of the art segmentation techniques that are generally used for processing biological images. 

\section{Method: Coupled TuFF-BFF}
\label{method}
A flow-field technique is an approach to segmentation that uses a vector to extend the segmented region. Coupled TuFF-BFF is an automatic microglia segmentation algorithm that optimally combines the tubularity flow field technique (TuFF) \cite{mukherjee2015tubularity} with a blob flow field (BFF) technique. The TuFF algorithm is specific to neuron dendritic trees because it only searches for tubular structures in an image. The fine processes of microglia do have tubular shapes, but the TuFF algorithm does not account for the microglia soma. Our coupled TuFF-BFF algorithm segments both the processes and soma while minimizing the overlap of their segmentation. 

Coupled TuFF-BFF is in the family of active contour models that pull a contour or snake towards the edges or lines of the object in an image \cite{kass1988snakes, malladi1995shape, li2007active, mansouri2004constraining, ray2002active, goobic2005image, cui2006monte}. The snake is evolved by minimizing an energy functional, $\varepsilon(\phi)$, that follows some constraints until it converges to the object boundary, or zero level set. $\phi$ is the level set function that is positive inside the zero level set and negative on the outside. 

\subsection{Tubular Flow Field algorithm}
TuFF \cite{mukherjee2015tubularity} uses the tubular structure of the vessel-like objects to evolve a level set towards the objects boundary. The evolution of the contour relies on the tubular vector field of the image \cite{li2007active} which is attained by the orthonormal eigenvectors $\textbf{e}_i(\textbf{x})$, where \textbf{x} is the pixel position within the image domain $\Omega$. The eigenvectors are ordered by increasing magnitude of the eigenvalues, $|\lambda_1| \leq |\lambda_2| \leq |\lambda_3| >>0$. These eigenvalues are attained by computing the Hessian matrix of the Gaussian-smoothed image. The algorithm uses Frangi's vessel enhancement technique\cite{frangi1998multiscale} to distinguish and enhance tubular structures in an image by using a multiscale vesselness function according to the three directions of $\textbf{e}_i(\textbf{x})$. The segmentation is achieved by minimizing an energy functional $\varepsilon(\phi)$:
\vspace{-0.2cm}
\begin{equation}
\varepsilon(\phi)=\varepsilon_{reg}(\phi)+\varepsilon_{evolve}(\phi)+\varepsilon_{attr}(\phi)
\end{equation}
\label{eq:energy}

\vspace{-1cm}
\begin{equation}
\varepsilon_{reg}(\phi)= v_1\int_{\Omega} |\nabla \textit{H}(\phi)|\textit{d}\textbf{x}
\end{equation}
\label{eq:energyreg}
\vspace{-1cm}

\begin{equation}
\varepsilon_{evolve}(\phi)= -\int_{\Omega}\sum_{i=1}^d \alpha_i(\textbf{x})\langle\textbf{e}_i(\textbf{x}),\textbf{n}(\textbf{x})\rangle^2 \textit{H}(\phi)\textit{d}\textbf{x}
\end{equation}
\label{eq:energyevolve}
\vspace{-0.2cm}

where $\varepsilon(x)_{reg}$ is the smoothness energy, $\varepsilon(x)_{evolve}$ is the curve evolution energy, and $\varepsilon(x)_{attr}$ is the attraction energy. The smoothness weight, $v_1$, controls the smoothness of the level set curve. $\varepsilon(x)_{reg}$ constrains the length of the zero level set with the gradient of the Heaviside function in terms of $\phi$. The vector $\textbf{n}(\textbf{x}) $ is the outward normal to the zero level set of $\phi$ which effects the evolution along the vessel width. $\varepsilon_{attr}(\phi)$ is the attractive energy which uses the vector field to connect smaller disjoint fragments to larger fragments during the segmentation. The energy functional, $\varepsilon(\phi)$, is minimized where $\phi$ is iteratively updated using gradient descent \cite{mukherjee2015tubularity}.

%\subsection{Blob Flow Field algorithm}
%Similar to the tubularity measure, our method uses a blobness vector field to the algorithm to account for the soma of the cell. The blobness measure is calculated by again ordering the eigenvalues of the Hessian matrix by increasing magnitudes, $|\lambda_1| \leq |\lambda_2| \leq |\lambda_3|$, to attain a structure that has high magnitude of $\lambda$ in three orthonormal directions, thus the three eigenvalues are high, $|\lambda_i| >>0$ \cite{frangi1998multiscale, antiga2007generalizing}. 

\subsection{Coupled TuFF-BFF for reconstruction of microglia}
Similar to the tubularity measure, the proposed method uses a blobness vector field in the algorithm to account for the soma of the cell. Since the soma and the processes have varying thickness, we scale the width of the Gaussian corresponding to their sizes, where the width of the soma is to be much larger than the width of the fine processes. The blobness measure is calculated by again ordering the eigenvalues of the Hessian matrix by increasing magnitudes, $|\lambda_1| \leq |\lambda_2| \leq |\lambda_3|$ to attain a structure that has high magnitude of $\lambda$ in three orthonormal directions \cite{frangi1998multiscale, antiga2007generalizing}.
\begin{figure*}[t!]
	\centering
	\renewcommand{\tabcolsep}{0.05cm}
	\setlength{\belowcaptionskip}{-10pt}
	\begin{adjustbox}{width=.95\textwidth}
	{
			\begin{tabular}{ccccc}
				%	\begin{tabular}{@{}p{2cm}p{2cm}p{2cm}p{2cm}p{2cm}p{2cm}p{2cm}p{2cm}p{2cm}p{2cm}@{}}
				{Original} & {Ground truth} & {Coupled TuFF-BFF} & {L2S\cite{mukherjee2015region}} & {Chan-Vese\cite{chan2001active}}
				\\
				\includegraphics[width=.16\linewidth, height = 0.12\linewidth, scale=0.1]{images/orig_maygreat.png} &
				\includegraphics[width=.16\linewidth, height = 0.12\linewidth, scale=0.1]{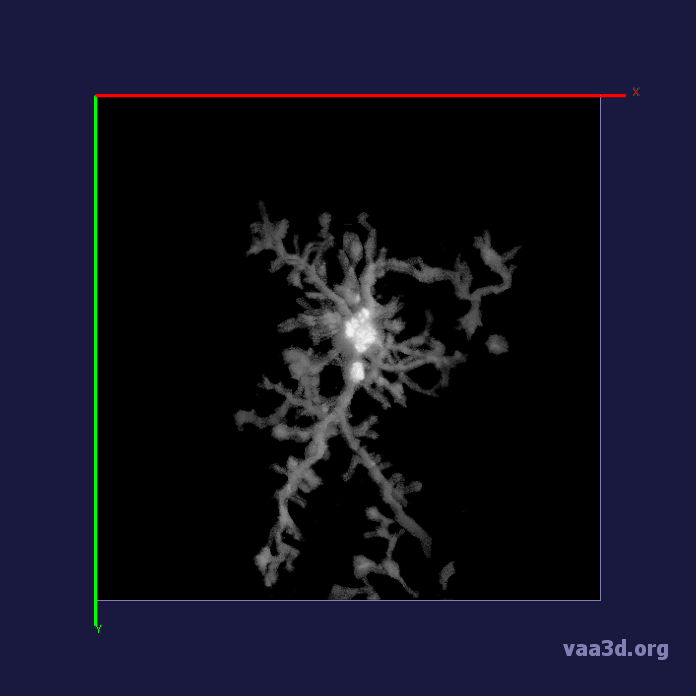} &
				\includegraphics[width=.16\linewidth, height = 0.12\linewidth, scale=0.1]{images/bff_maygreat.png} &
				\includegraphics[width=.16\linewidth, height = 0.12\linewidth, scale=0.1]{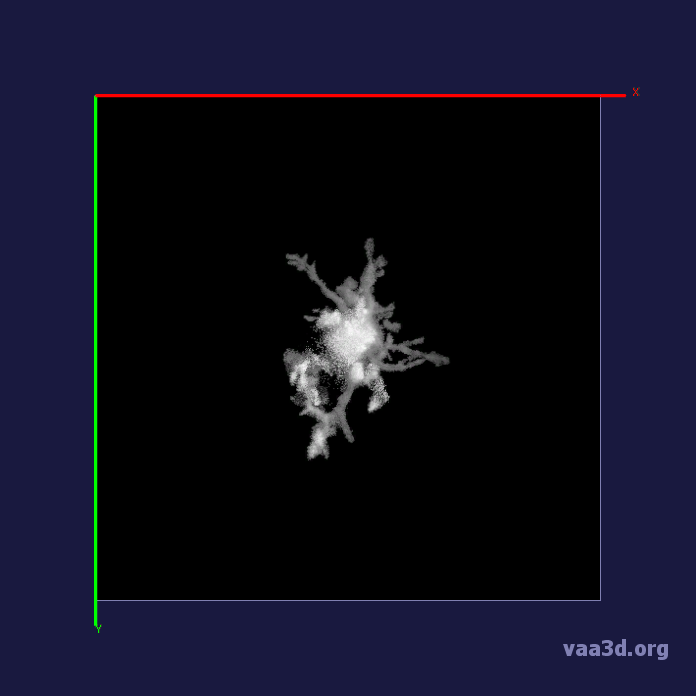} &
				\includegraphics[width=.16\linewidth, height = 0.12\linewidth, scale=0.1]{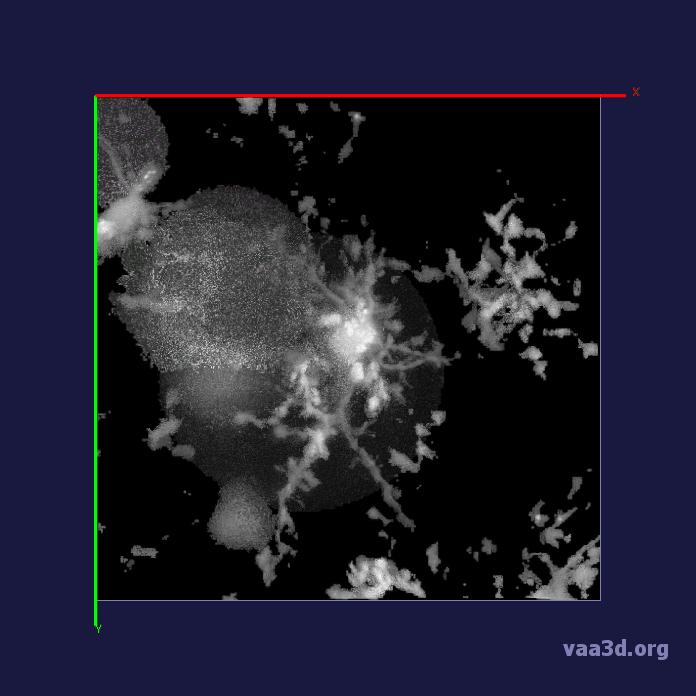} 
					\\
					\includegraphics[width=.16\linewidth, height = 0.12\linewidth, scale=0.1]{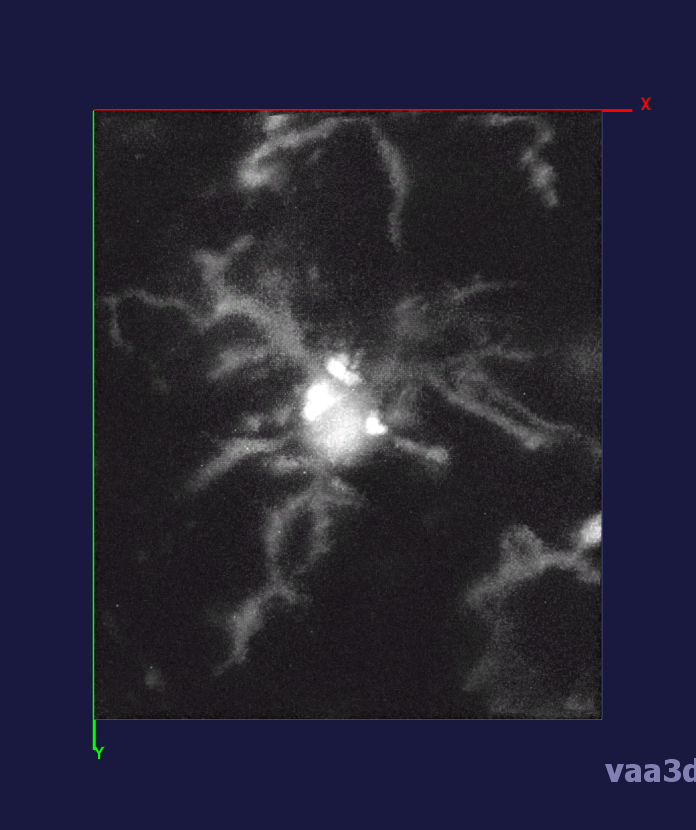} &
				\includegraphics[width=.16\linewidth, height = 0.12\linewidth, scale=0.1]{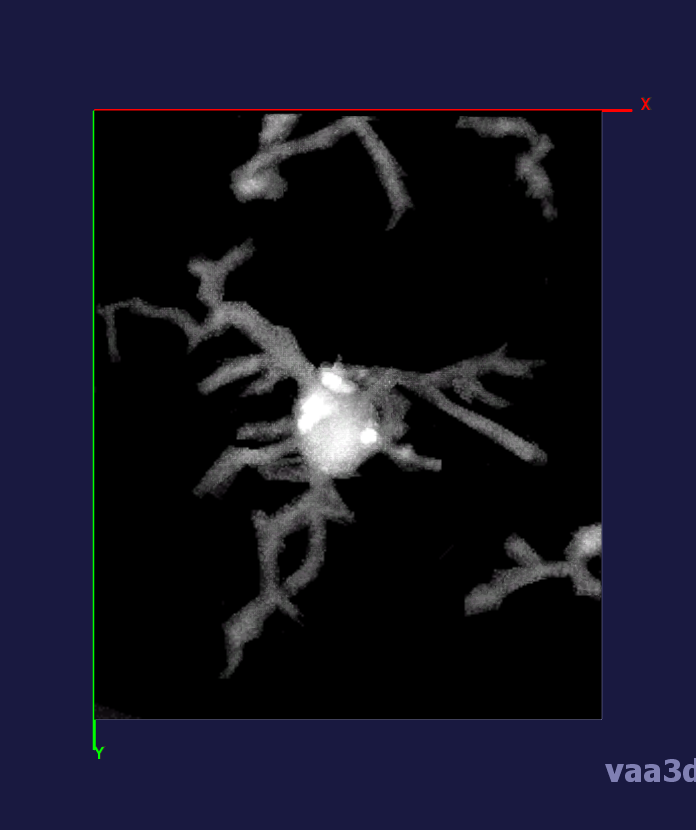} &
				\includegraphics[width=.16\linewidth, height = 0.12\linewidth, scale=0.1]{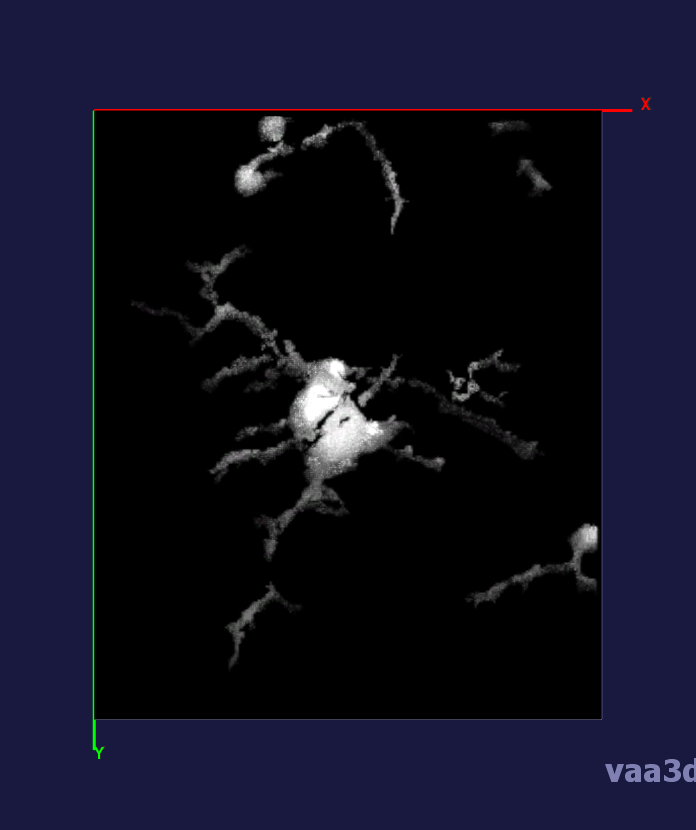} &
				\includegraphics[width=.16\linewidth, height = 0.12\linewidth, scale=0.1]{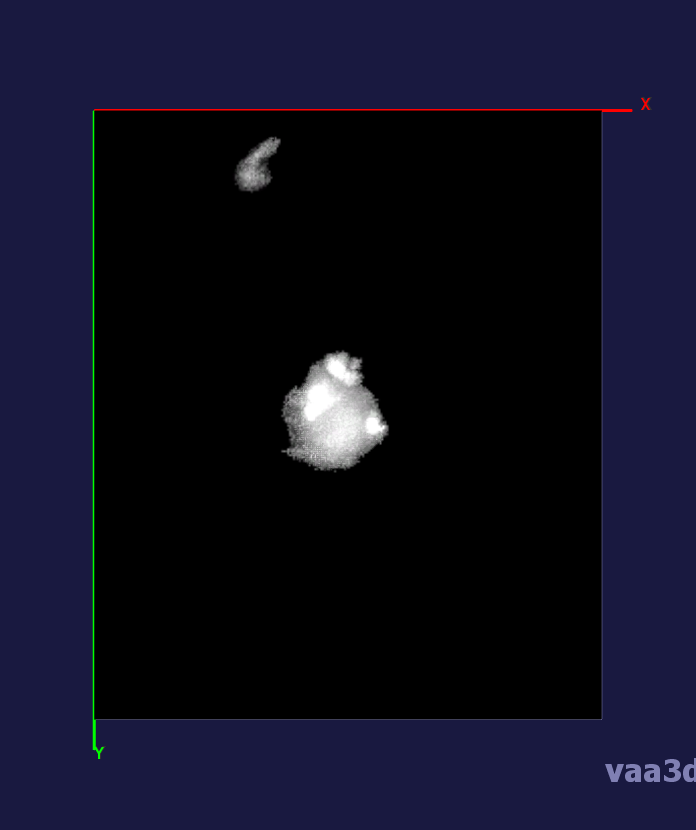} &
				\includegraphics[width=.16\linewidth, height = 0.12\linewidth, scale=0.1]{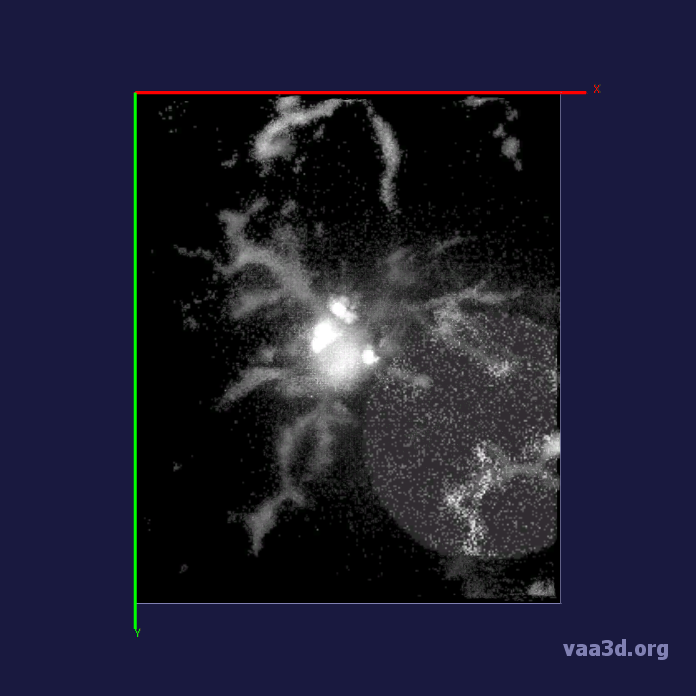} 
				\\
					
					\includegraphics[width=.16\linewidth, height = 0.12\linewidth, scale=0.1]{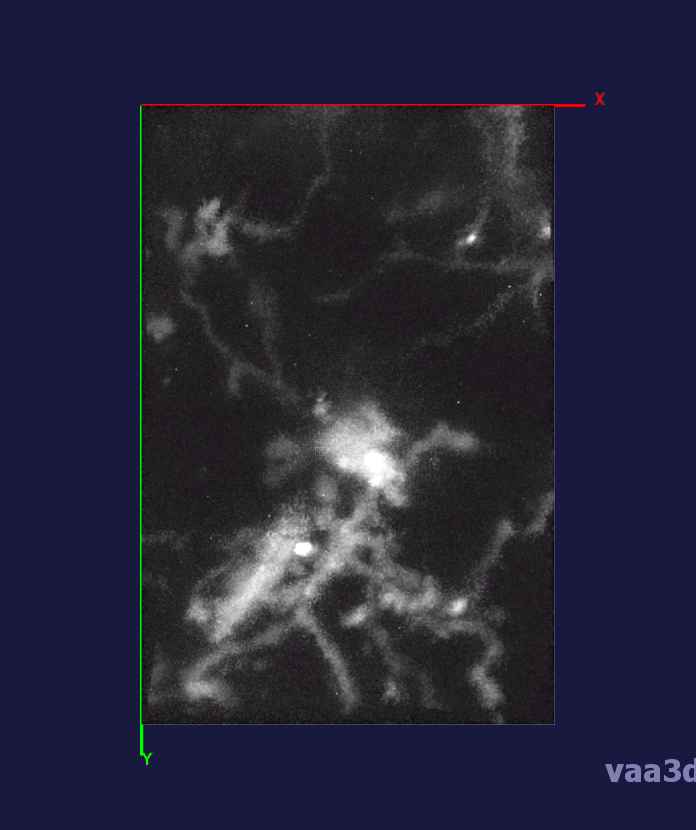} &
				\includegraphics[width=.16\linewidth, height = 0.12\linewidth, scale=0.1]{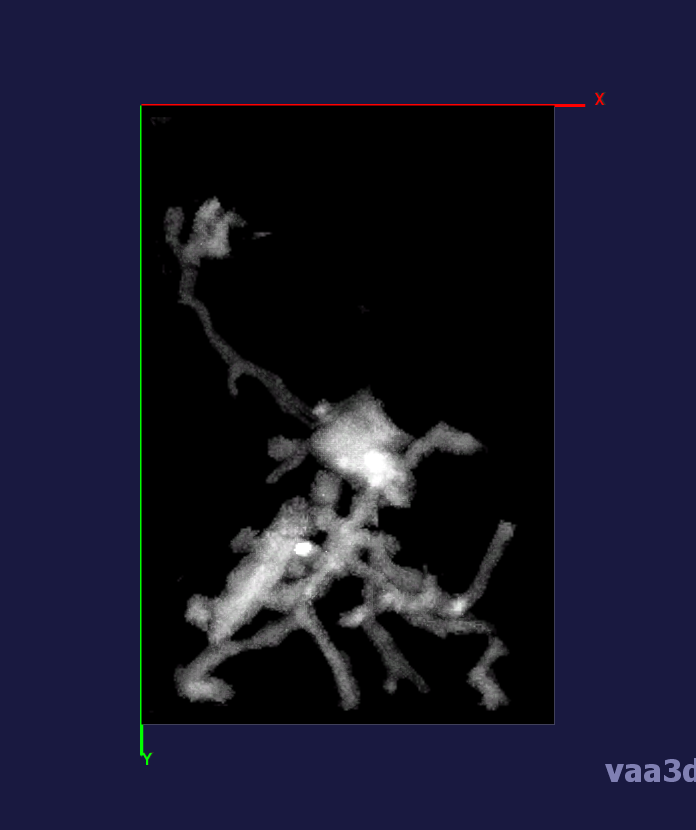} &
				\includegraphics[width=.16\linewidth, height = 0.12\linewidth, scale=0.1]{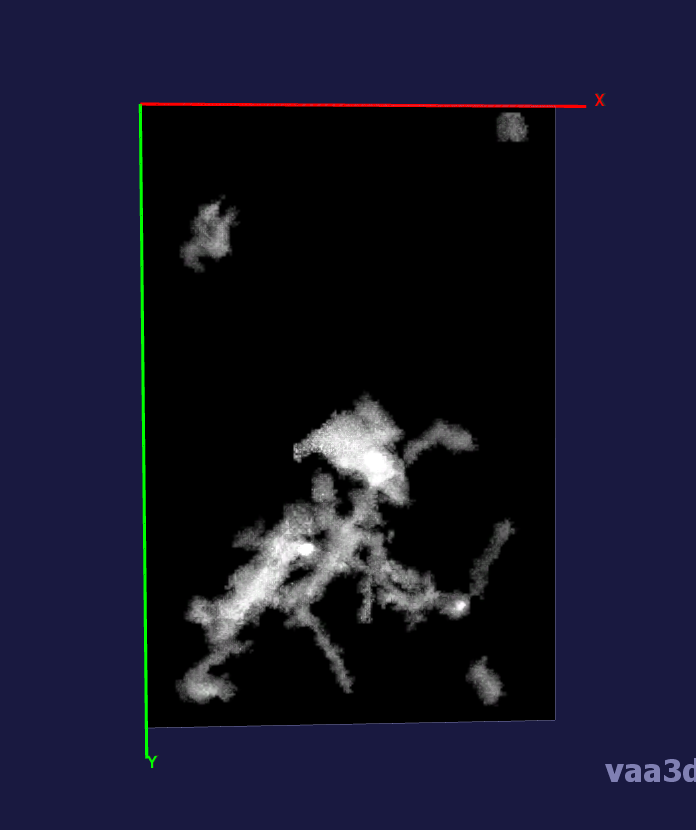} &
				\includegraphics[width=.16\linewidth, height = 0.12\linewidth, scale=0.1]{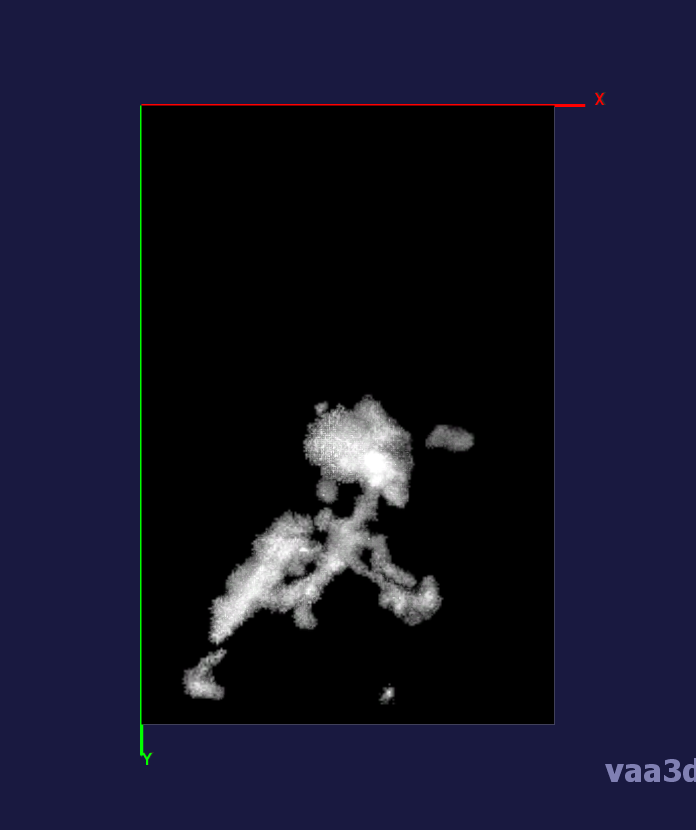} &
				\includegraphics[width=.16\linewidth, height = 0.12\linewidth, scale=0.1]{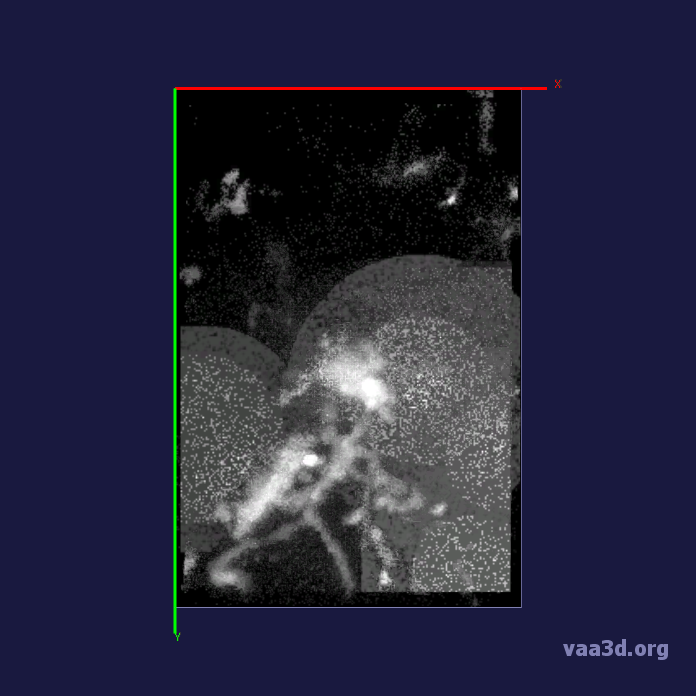} 
					\\
					
					\includegraphics[width=.16\linewidth, height = 0.12\linewidth, scale=0.1]{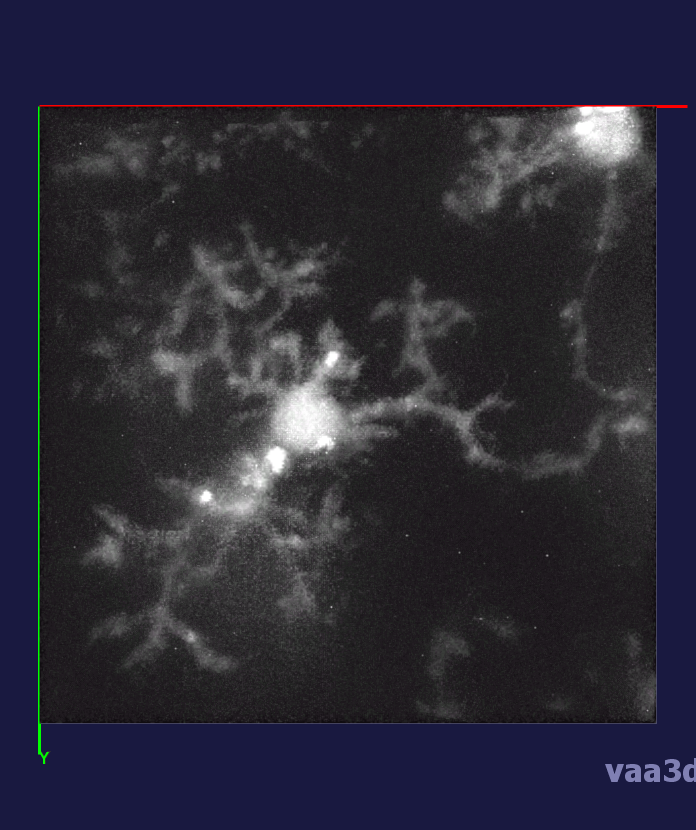} &
				\includegraphics[width=.16\linewidth, height = 0.12\linewidth, scale=0.1]{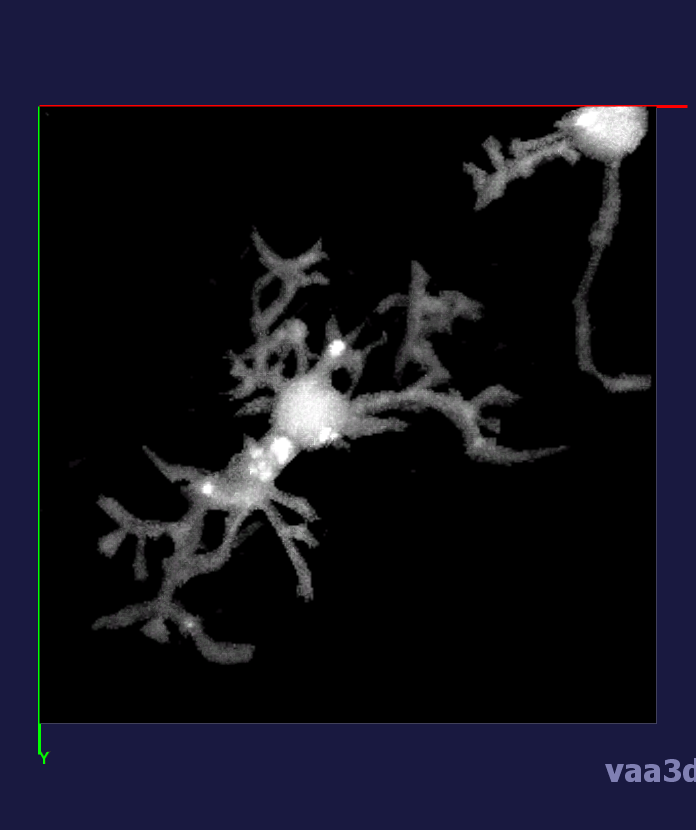} &
				\includegraphics[width=.16\linewidth, height = 0.12\linewidth, scale=0.1]{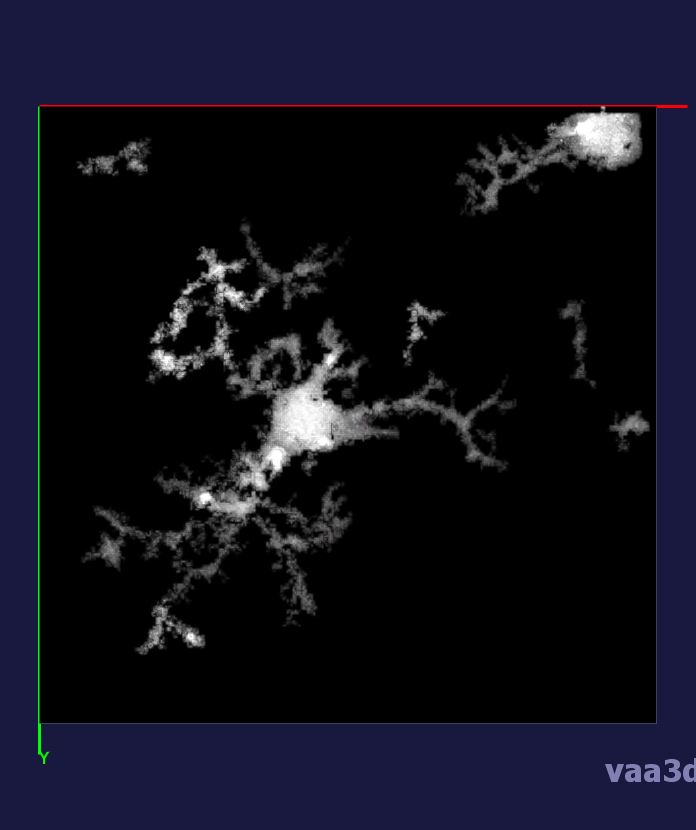} &
				\includegraphics[width=.16\linewidth, height = 0.12\linewidth, scale=0.1]{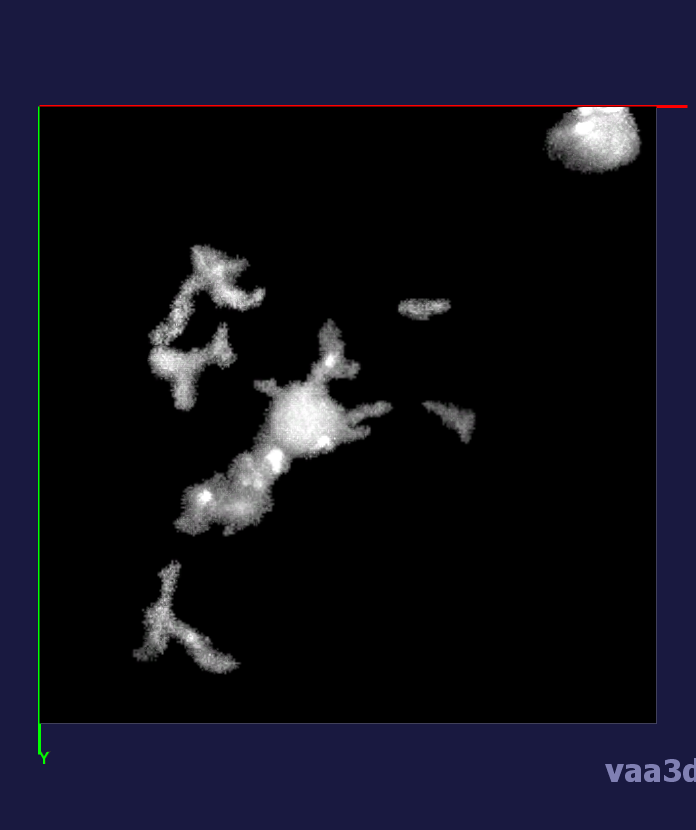} &
				\includegraphics[width=.16\linewidth, height = 0.12\linewidth, scale=0.1]{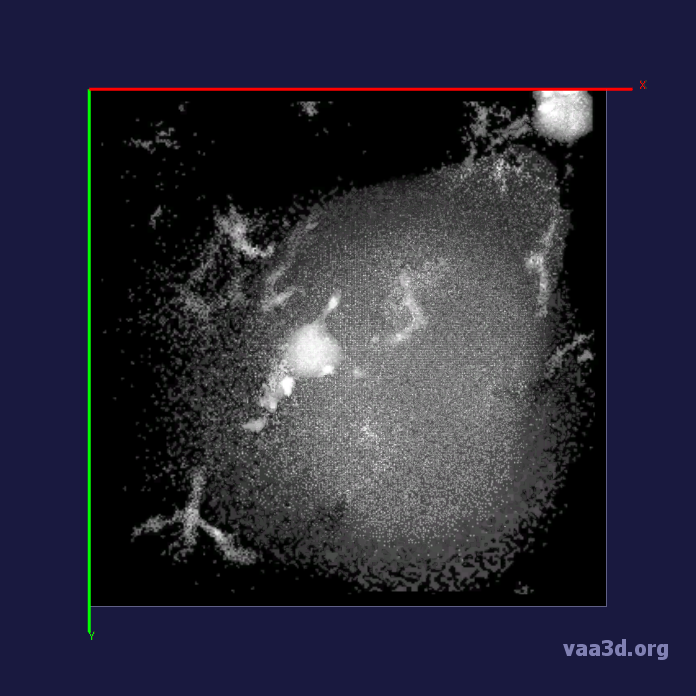} 
%				
%				\\
%					
%					\includegraphics[width=.18\linewidth, height = 0.13\linewidth, scale=0.1]{images/orig_may750_1000.png} &
%				\includegraphics[width=.18\linewidth, height = 0.13\linewidth, scale=0.1]{images/gt_may750_1000.png} &
%				\includegraphics[width=.18\linewidth, height = 0.13\linewidth, scale=0.1]{images/bff_may750_1000.png} &
%				\includegraphics[width=.18\linewidth, height = 0.13\linewidth, scale=0.1]{images/l2s_may750_1000.png} &
%				\includegraphics[width=.18\linewidth, height = 0.13\linewidth, scale=0.1]{images/orig_may750_1000.png} 
%				
%			
			\end{tabular}
		}	
	\end{adjustbox}
%	\vspace{-0.1cm}
	\caption{\small{Segmentation results of 3D microglia images.}}
	
	\vspace{-0.5cm}
	\label{fig:results}
\end{figure*}

After computing the tubular and blobness information, the initial level set is attained from the 3D stack. The level set contours $\phi_1$ to capture the processes and $\phi_2$ to capture the soma are separately initialized by Otsu thresholding \cite{otsu1979threshold} the image's vessel- and blob-enhanced image. The processes and soma of microglia are simultaneously segmented by evolving their level sets and minimizing their respective energy functionals, $\varepsilon_{TuFF}(\phi_1)$ and $\varepsilon_{BFF}(\phi_2)$:
\vspace{-0.2cm}
\begin{equation}
\varepsilon_{TuFF}(\phi_1)=\varepsilon_{reg}(\phi_1)+\varepsilon_{evolve}(\phi_1)+\varepsilon_{attr}(\phi_1)+\varepsilon_{repel}(\phi_2)
\end{equation}
\label{eq:energytuff}

\vspace{-1cm}
\begin{equation}
\varepsilon_{BFF}(\phi_2)=\varepsilon_{reg}(\phi_2)+\varepsilon_{evolve}(\phi_2)+\varepsilon_{attr}(\phi_2)+\varepsilon_{repel}(\phi_1)
\end{equation}
\label{eq:energybff}

\vspace{-1cm}
\begin{equation}
\varepsilon_{repel}(\phi_{i})=\int_\Omega \textit{H} (\phi_{TuFF})\textit{H} (\phi_{BFF})dx
\end{equation}
\label{eq:repel}
Although the vesselness and blobness segmentations are separate, they are linked by using the result of both level sets in the $\varepsilon_{repel}(\phi)$ term. $\varepsilon_{repel}(\phi_{i})$ penalizes the regions of overlap between the two level sets. The level set functions $\phi$ can be iteratively updated by solving $\frac{\partial \varepsilon}{\partial \phi}$ which, by the chain rule, can be solved with $ \frac{\partial \phi}{\partial t}$, where \textit{t} denotes each iteration\cite{mukherjee2015tubularity}. We call this \textit{F}, the velocity of the level set implementation: 
\begin{equation}
F = \frac{\partial \phi_{reg}}{\partial t}+ \alpha \frac{\partial \phi_{evolve}}{\partial t} + v_1 \frac{\partial \phi_{attr}}{\partial t} + \frac{r \partial \phi_{repel}}{\partial t}
\end{equation}\label{eq:bffvel}
The regions of overlap between both level sets are computed for $\frac{r \partial \phi_{repel}}{\partial t}$, where the repel term $r=0$ when there is no overlap. This term changes the velocity, \textit{F}, within the overlapping regions to repel away from their opposing level set $\phi$. Thus, the repel force energy functional $\varepsilon_{repel}(\phi)$ minimizes the overlap between the segmentation of the processes and soma to attain a joint segmentation. 

\vspace{-.3cm}
\section{Experimental Results and Analysis}
\label{sec: exp}
%explain images, imaging technique (including flourescence and microscopy), noise, intensity variation 
The dataset consists of 3D images of microglia imaged from healthy mice brains using multiphoton microscopy.

\begin{figure}
	\centering
  \includegraphics[width=.8\linewidth]{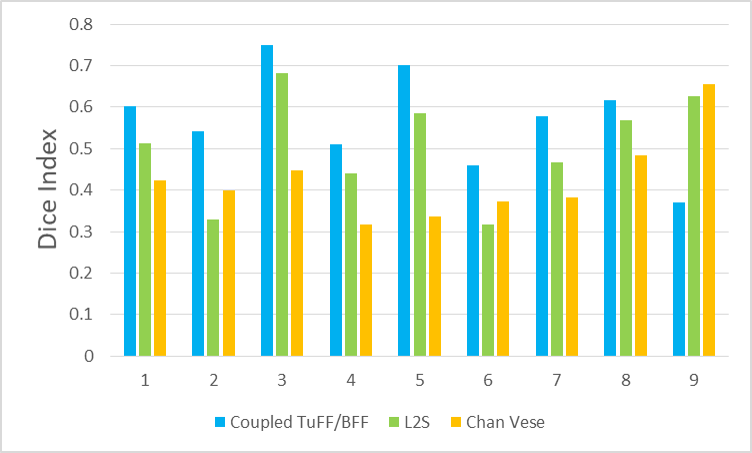}
  \caption{\small{Dice index of the segmentation using Coupled TuFF-BFF, L2S\cite{mukherjee2015region}, and Chan-Vese\cite{chan2001active}.}}
  \label{fig:dice}
  	\vspace{-0.5cm}
\end{figure} 
\vspace{-.5cm}
\subsection{Imaging and fluorescence technique}
The dataset consists of 3D images of microglia from mice using multiphoton microscopy. To label microglia in the mouse brain we used mice with an inducible cre recombinase under the control of the CX3CR1 promoter crossed to the Ai6 fluorescent reporter mouse  (Jackson Laboratories, Bar Harbor, ME) to generate CX3CR1creERT2/+ X Ai6ZsGreen \cite{yona2013fate, madisen2010robust}. At post-natal day (P23) 23, mice were given 10uL/g body weight of a 20mg/mL Tamoxifen (Sigma) solution in corn oil to induce recombination of the floxed stop codon leading to ZsGreen expression in microglia. All procedures adhered to guidelines of the Institutional Animal Care and Use Committee (ACUC) at the University of Virginia. Microglia of adult mice (7-10 weeks old) were imaged using a Leica TCS SP8 multiphoton microscopy system equipped with a equipped with a Coherent Chameleon Ti:Sapphire laser and a 25x 0.95 NA immersion lens. ZsGreen was excited with a wavelength of 880 nm.
 \vspace{-.3cm}
\subsection{Dataset}
The 3D movies of microglia were imaged over 20 minutes with z-stacks taken at one minute intervals, containing single or multiple microglia per field of view. Some of the images were cropped from a larger field of view containing about 10 different cells and two images were imaged from a zoomed in view of one individual cell. The images ranged from a horizontal pixel width of .01 um and a vertical pixel width of .01 um to horizontal pixel width of .2 um and a vertical pixel width of .2 um. In the 3D images, there is variation in intensity contrast throughout the cell, non-structural noise, and fluorescence bleeding through z-stack due to  the lengthy imaging technique which makes it difficult to visualize and process. The images were pre-processed using histogram equalization which increased the intensity throughout the cell but further increased noise in the background. 

The parameter for the width of the Gaussian filter is dependent on imaging depth. For our experiments we used $\sigma=$ 0.5 to 1 to find the processes and $\sigma=$ 4 to 7 to attain the soma structure. The smoothness parameter was set from $v_1 =$ .02 to .09 to attain the best segmentation results. All experiments required fewer than 50 iterations.

\begin{figure}
	\centering
  \includegraphics[width=.9\linewidth]{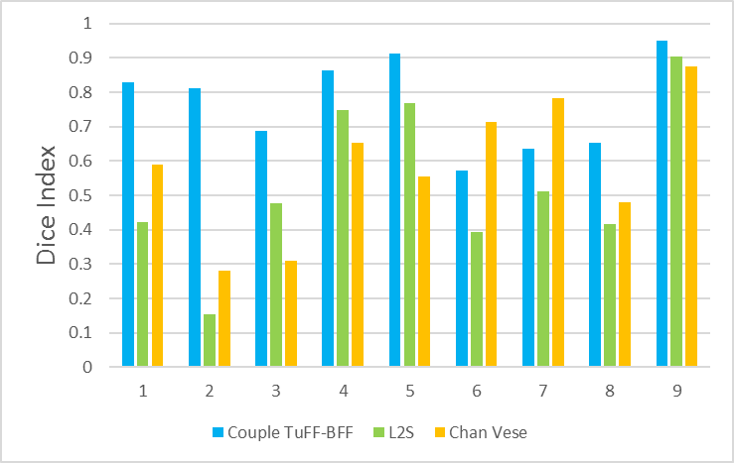}
  \caption{\small{Dice index of surveyed area from the segmentation using Coupled TuFF-BFF, L2S\cite{mukherjee2015region}, and Chan-Vese\cite{chan2001active}.}}
  \label{fig:diceCH}
 \vspace{-.5cm}
  \end{figure}
  
\subsection{Performance evaluation}
In our experiments, we compare the coupled TuFF-BFF microglia segmentation results with those given by L2S \cite{mukherjee2015region} and the Chan-Vese segmentation method \cite{chan2001active}. The groundtruth in 3D was attained by manually tracing the object slice by slice from the z-stack. It must be noted that this was done by eye and could have some error. Figure \ref{fig:results} shows the visual comparison of the segmentation results for our dataset. Our result shown on the third column captures both the soma and processes. Figure \ref{fig:dice} shows the Dice coefficient comparison of each segmentation method to the ground truth. Since the soma is much larger than the fine processes in the microglia, the processes have less volumetric impact on the similarity score. As explained in Section \ref{sec:intro}, segmenting the processes is important for quantifying the extension from the soma and its volume of surveillance. We use the Dice coefficient to quantitatively compare the ramification by taking the convex hull of the resulting segmentation. The Dice coefficient is a similarity measure that is computed using with $2*\frac{|intersection(A,B)|}{(|A| + |B|)}$ where $A$ is the ground truth and $B$ is the compared image. 

From Figure \ref{fig:diceCH}, the average Dice score for coupled TuFF-BFF was 0.77, compared to 0.53 for L2S \cite{mukherjee2015region} and .58 for Chan-Vese \cite{chan2001active}. It must be noted that L2S required manual user initialization for each 2D image in the stack. While the Chan-Vese method has automatic seed selection, our coupled TuFF/BFF method was the only method that was a true 3D segmentation algorithm. L2S could not consistently capture the entire processes due to the intensity inhomogeneity throughout the object and background noise. The Chan-Vese segmentation could capture the extensions of the processes but did not work well with noise and attained false positives in the reconstruction. Since our method uses the tubular and blob information of the object to separate foreground and background, the segmentation only evolved within the object boundaries. 

From the segmentation of microglia from 3D multiphoton images, we attained quantification of the ramification of the microglia processes using the index provided by Madry \textit{et al.} The ramification index in Table 1 quantifies the extension of the processes from the soma. The ramification index of 1 is the soma with no ramification and a larger index denotes greater ramification. We compare the ramification index attained from the segmentation result from each method with that attained from the ground truth. The mean absolute error for coupled TuFF-BFF was 1.49 compared with 3.92 and 3.78 for L2S \cite{mukherjee2015region} and Chan-Vese \cite{chan2001active}, respectively. 

 %L2S is a region based segmentation method is the state of the art in segmentation of biological images containing tubular structures and claimed to perform well on images with inhomogeneous intensity. L2S requires manual initialization by the user. 

%  \begin{figure}
%	\centering
%  \includegraphics[width=.8\linewidth]{images/RamificatioinIndex.png}
%  \caption{\small{Ramification Index}}
%  \label{fig:diceCH}
%    \end{figure}

  \begin{center}
  \textbf{Table 1} \ \  Ramification Index\\
 
\begin{tabular}{ccccc} 
%\label{table: ram}
\hline
{No.} & groundtruth & TuFF-BFF& L2S    & Chan-Vese\\ \hline
 \#1 & 8.88 & 7.88 & 4.0 & 7.46 \\ 
 \#2 & 7.69 & 10.14 & 2.1 & 9.89 \\ 
 \#3 & 6.54 & 5.98 & 4.34 & 8.76\\ 
 \#4 & 9.02 & 13.6 & 5.48 & 12.4\\ 
 \#5 & 6.44 & 7.22 & 5.26 & 18.3\\ 
 \#6 &  8.60 & 8.74 & 3.57 & 11.0\\ 
 \#7 & 9.09 & 7.70 & 4.78 & 12.86\\ 
 \#8 & 8.88 & 7.88 & 4.0 & 7.46\\ 
 \#9 & 11.18 & 12.7 & 7.56 & 16.48\\ \hline 
{MAE:} & -- & 1.49 & 3.92 & 3.78\\
\hline \\
   \label{table: ram}
\end{tabular}
\vspace{-0.5cm}
\end{center}
 % \begin{figure}
 % \centering
 % \includegraphics[width=.9\linewidth]{images/CumSurv.png}
  %\caption{\small{Cumulative Surveillance .}}
 % \label{fig:cumsurv}
 
 % \end{figure}
\section{Conclusion}
In this paper, we proposed an automated segmentation method that captured microglia in 3D images. There was no smoothing or enhancement to the image prior to the application of our algorithm. Coupled TuFF-BFF was able to segment processes and soma from 3D images of microglia from the mouse brain. It was able to simultaneously capture the object of interest from images despite intensity inhomogeneity throughout the cell and background noise. While our method performed better than the state of the art, it could be further improved to attain a more accurate thickness of the cell and capture the low intensity areas of the branches. We plan to apply our method on images of microglia from mice in other states that significantly alters the microglia morphology. Another extension planned involves using coupled TuFF-BFF to extending existing cell tracking algorithms \cite{mansouri2004constraining, goobic2005image, cui2006monte}

\clearpage

\bibliographystyle{IEEEtran}
{\small
\bibliography{BFF}}

\begin{thebibliography}{10}
\providecommand{\url}[1]{#1}
\csname url@samestyle\endcsname
\providecommand{\newblock}{\relax}
\providecommand{\bibinfo}[2]{#2}
\providecommand{\BIBentrySTDinterwordspacing}{\spaceskip=0pt\relax}
\providecommand{\BIBentryALTinterwordstretchfactor}{4}
\providecommand{\BIBentryALTinterwordspacing}{\spaceskip=\fontdimen2\font plus
\BIBentryALTinterwordstretchfactor\fontdimen3\font minus
  \fontdimen4\font\relax}
\providecommand{\BIBforeignlanguage}[2]{{%
\expandafter\ifx\csname l@#1\endcsname\relax
\typeout{** WARNING: IEEEtran.bst: No hyphenation pattern has been}%
\typeout{** loaded for the language `#1'. Using the pattern for}%
\typeout{** the default language instead.}%
\else
\language=\csname l@#1\endcsname
\fi
#2}}
\providecommand{\BIBdecl}{\relax}
\BIBdecl

\bibitem{colonna2017microglia}
M.~Colonna and O.~Butovsky, ``Microglia function in the central nervous system
  during health and neurodegeneration,'' \emph{Annual Review of Immunology},
  vol.~35, pp. 441--468, 2017.

\bibitem{schafer2015microglia}
D.~P. Schafer and B.~Stevens, ``Microglia function in central nervous system
  development and plasticity,'' \emph{Cold Spring Harbor Perspectives in
  Biology}, vol.~7, no.~10, p. a020545, 2015.

\bibitem{nimmerjahn2005resting}
A.~Nimmerjahn, F.~Kirchhoff, and F.~Helmchen, ``Resting microglial cells are
  highly dynamic surveillants of brain parenchyma in vivo,'' \emph{Science},
  vol. 308, no. 5726, pp. 1314--1318, 2005.

\bibitem{davalos2012fibrinogen}
D.~Davalos, J.~K. Ryu, M.~Merlini, K.~M. Baeten, N.~Le~Moan, M.~A. Petersen,
  T.~J. Deerinck, D.~S. Smirnoff, C.~Bedard, H.~Hakozaki \emph{et~al.},
  ``Fibrinogen-induced perivascular microglial clustering is required for the
  development of axonal damage in neuroinflammation,'' \emph{Nature
  Communications}, vol.~3, p. 1227, 2012.

\bibitem{perry2010microglia}
V.~H. Perry, J.~A. Nicoll, and C.~Holmes, ``Microglia in neurodegenerative
  disease,'' \emph{Nature Reviews Neurology}, vol.~6, no.~4, p. 193, 2010.

\bibitem{wake2009resting}
H.~Wake, A.~J. Moorhouse, S.~Jinno, S.~Kohsaka, and J.~Nabekura, ``Resting
  microglia directly monitor the functional state of synapses in vivo and
  determine the fate of ischemic terminals,'' \emph{Journal of Neuroscience},
  vol.~29, no.~13, pp. 3974--3980, 2009.

\bibitem{tremblay2010microglial}
M.-{\`E}. Tremblay, R.~L. Lowery, and A.~K. Majewska, ``Microglial interactions
  with synapses are modulated by visual experience,'' \emph{PLoS Biology},
  vol.~8, no.~11, p. e1000527, 2010.

\bibitem{madry2017microglial}
C.~Madry, V.~Kyrargyri, I.~L. Arancibia-C{\'a}rcamo, R.~Jolivet, S.~Kohsaka,
  R.~M. Bryan, and D.~Attwell, ``Microglial ramification, surveillance, and
  interleukin-1$\beta$ release are regulated by the two-pore domain k+ channel
  thik-1,'' \emph{Neuron}, 2017.

\bibitem{davalos2005atp}
D.~Davalos, J.~Grutzendler, G.~Yang, J.~V. Kim, Y.~Zuo, S.~Jung, D.~R. Littman,
  M.~L. Dustin, and W.-B. Gan, ``Atp mediates rapid microglial response to
  local brain injury in vivo,'' \emph{Nature Neuroscience}, vol.~8, no.~6, p.
  752, 2005.

\bibitem{gyoneva2014systemic}
S.~Gyoneva, D.~Davalos, D.~Biswas, S.~A. Swanger, E.~Garnier-Amblard, F.~Loth,
  K.~Akassoglou, and S.~F. Traynelis, ``Systemic inflammation regulates
  microglial responses to tissue damage in vivo,'' \emph{Glia}, vol.~62, no.~8,
  pp. 1345--1360, 2014.

\bibitem{mukherjee2015tubularity}
S.~Mukherjee, B.~Condron, and S.~T. Acton, ``Tubularity flow field — a
  technique for automatic neuron segmentation,'' \emph{IEEE Transactions on
  Image Processing}, vol.~24, no.~1, pp. 374--389, 2015.

\bibitem{kass1988snakes}
M.~Kass, A.~Witkin, and D.~Terzopoulos, ``Snakes: Active contour models,''
  \emph{International Journal of Computer Vision}, vol.~1, no.~4, pp. 321--331,
  1988.

\bibitem{malladi1995shape}
R.~Malladi, J.~A. Sethian, and B.~C. Vemuri, ``Shape modeling with front
  propagation: A level set approach,'' \emph{IEEE Transactions on Pattern
  Analysis and Machine Intelligence}, vol.~17, no.~2, pp. 158--175, 1995.

\bibitem{li2007active}
B.~Li and S.~T. Acton, ``Active contour external force using vector field
  convolution for image segmentation,'' \emph{IEEE Transactions on Image
  Processing}, vol.~16, no.~8, pp. 2096--2106, 2007.

\bibitem{mansouri2004constraining}
A.-R. Mansouri, D.~P. Mukherjee, and S.~T. Acton, ``Constraining active contour
  evolution via lie groups of transformation,'' \emph{IEEE Transactions on
  Image Processing}, vol.~13, no.~6, pp. 853--863, 2004.

\bibitem{ray2002active}
N.~Ray and S.~T. Acton, ``Active contours for cell tracking,'' in \emph{Image
  Analysis and Interpretation, 2002. Proceedings. Fifth IEEE Southwest
  Symposium on}.\hskip 1em plus 0.5em minus 0.4em\relax IEEE, 2002, pp.
  274--278.

\bibitem{goobic2005image}
A.~P. Goobic, J.~Tang, and S.~T. Acton, ``Image stabilization and registration
  for tracking cells in the microvasculature,'' \emph{IEEE Transactions on
  Biomedical Engineering}, vol.~52, no.~2, pp. 287--299, 2005.

\bibitem{cui2006monte}
J.~Cui, S.~T. Acton, and Z.~Lin, ``A monte carlo approach to rolling leukocyte
  tracking in vivo,'' \emph{Medical Image Analysis}, vol.~10, no.~4, pp.
  598--610, 2006.

\bibitem{frangi1998multiscale}
A.~F. Frangi, W.~J. Niessen, K.~L. Vincken, and M.~A. Viergever, ``Multiscale
  vessel enhancement filtering,'' in \emph{International Conference on Medical
  Image Computing and Computer-Assisted Intervention}.\hskip 1em plus 0.5em
  minus 0.4em\relax Springer, 1998, pp. 130--137.

\bibitem{antiga2007generalizing}
L.~Antiga, ``Generalizing vesselness with respect to dimensionality and
  shape,'' \emph{The Insight Journal}, vol.~3, pp. 1--14, 2007.

\bibitem{mukherjee2015region}
S.~Mukherjee and S.~T. Acton, ``Region based segmentation in presence of
  intensity inhomogeneity using legendre polynomials,'' \emph{IEEE Signal
  Processing Letters}, vol.~22, no.~3, pp. 298--302, 2015.

\bibitem{chan2001active}
T.~F. Chan and L.~A. Vese, ``Active contours without edges,'' \emph{IEEE
  Transactions on Image Processing}, vol.~10, no.~2, pp. 266--277, 2001.

\bibitem{otsu1979threshold}
N.~Otsu, ``A threshold selection method from gray-level histograms,''
  \emph{IEEE Transactions on Systems, Man, and Cybernetics}, vol.~9, no.~1, pp.
  62--66, 1979.

\bibitem{yona2013fate}
S.~Yona, K.-W. Kim, Y.~Wolf, A.~Mildner, D.~Varol, M.~Breker, D.~Strauss-Ayali,
  S.~Viukov, M.~Guilliams, A.~Misharin \emph{et~al.}, ``Fate mapping reveals
  origins and dynamics of monocytes and tissue macrophages under homeostasis,''
  \emph{Immunity}, vol.~38, no.~1, pp. 79--91, 2013.

\bibitem{madisen2010robust}
L.~Madisen, T.~A. Zwingman, S.~M. Sunkin, S.~W. Oh, H.~A. Zariwala, H.~Gu,
  L.~L. Ng, R.~D. Palmiter, M.~J. Hawrylycz, A.~R. Jones \emph{et~al.}, ``A
  robust and high-throughput cre reporting and characterization system for the
  whole mouse brain,'' \emph{Nature Neuroscience}, vol.~13, no.~1, p. 133,
  2010.

\end{thebibliography}

\end{document}